\documentstyle[mprocl,epsf,epsfig]{article}

\begin{document}

\title{Chaotic exit to inflation:  the dynamics of pre-inflationary universes}

\author{H. P. de Oliveira}

\address{Universidade do Estado do Rio de Janeiro, Instituto de F\'\i sica\\
R. S\~ao Francisco Xavier, 524, CEP 20550-013, Rio de Janeiro, RJ, Brazil}

\author{I. Dami\~ao Soares }

\address{Centro Brasileiro de Pesquisas F\'\i sicas\\ Rua Dr. 
Xavier Sigaud, 150, CEP 22290, Rio de Janeiro, RJ, Brazil}

\author{T. J. Stuchi}
\address{Universidade Federal do Rio de Janeiro, Instituto de F\'\i sica\\
Caixa Postal 68528, CEP 21945-970, Rio de Janeiro, RJ, Brazil}

\maketitle\abstracts{We show that anisotropic Bianchi type-IX models, with
matter and cosmological constant have chaotic dynamics, connected to the
presence of a saddle-center in phase space. The topology of cylinders
emanating from unstable periodic orbits about the saddle-center provides an
invariant characterization of chaos in the models. The model can be thought to
describe the early stages of inflation, the way out to inflation being
chaotic.} 

Some important features in inflationary dynamics arise whenever anisotropy is
present even in the form of small perturbations. This point has not been
emphazised yet in the literature of inflation. The conjunction of the
cosmological constant, anisotropy and matter fields implies the existence of a
saddle-center $E$ in phase space\cite{sc}. As consequence we have a complex
dynamics, based on structures as homoclinic cylinders, which emanate from
unstable periodic orbits in the neighborhood of $E$. These
cylinders will cross each other transversally producing a chaotic dynamics,
analogous to the breaking of homoclinic curves in Poincar\'e homoclinic
phenomena\cite{ivano}. This structure constitutes an invariant
characterization of chaos. 

We consider anisotropic Bianchi IX cosmological models characterized by two
scale factors $A(t)$ and $B(t)$, with dust plus a cosmological constant
$\Lambda$. The dynamics of the models is governed by the Hamiltonian

\begin{equation}
H(A,B,P_A,P_B ) = \frac{P_A\,P_B}{4\,B} - \frac{A\,P_A^2}{8\,B^2} + 2\,A
- \frac{A^3}{2\,B^2}
 - 2\,\Lambda\,A\,B^2 - E_0 = 0,    
\end{equation}

\noindent where $P_A$ and $P_B$ are the momenta canonically conjugated to $A$
and $B$, respectively, and $E_0$ is a constant proportional to the total
energy of the models. The associated dynamical system has one saddle-center
$E$ in the finite region of the phase space with associated energy $E_0 =
E_{crit} = \frac{1}{\sqrt{4\,\Lambda}}$ and eigenvalues
$\lambda_{1,2}=\pm\,\frac{1}{2\,E_{crit}}$,
$\lambda_{3,4}=\pm{2\,i}{E_{crit}}$. The phase space has two critical points
at infinity, corresponding to the de Sitter solution one of them acting as an attractor. The dynamical system
generated by (1) admits the invariant manifold ${\cal{M}}$ ($A =B$, $P_A =
\frac{P_B}{2}$). Its phase portrait is depicted in Fig. 2, where the curves
represent isotropic universes. The critical point $E$ is contained in
${\cal{M}}$, and the separatrices $S$ constitute boundaries between
isotropic models that collapse and escape to the de Sitter configuration.

Due to its saddle-center character, the motion in a small neighborhood of the
critical point $E$ is separable into hyperbolic and rotational motions, with
respective energies approximately conserved. The rotational motion corresponds
to periodic orbits in a linear neighborhood of $E$ (cf. Fig 1(a)). For zero
energy of the hyperbolic motion, we have the linear stable $V_s$ and unstable
$V_u$ one-dimensional manifolds of Fig. 1(b). The separatrices $S$ in the
invariant manifold ${\cal{M}}$ are actually the non-linear extension of $V_s$
and $V_u$. The direct product of a periodic orbit with $V_s$ and $V_u$
generates in the linear neighborhood of $E$ the structure of stable and
unstable cylinders, which coalesce into the periodic orbit for large positive
and negative times, respectively (Fig. 1(c)). A general orbit which visits the
neighborhood of $E$ has an oscillatory approach to the cylinders, the closer
as the energy of the hyperbolic motion goes to zero. The outcome
of this oscillatory regime will be to collapse or to escape to the de Sitter
attractor depending on the energy of the hyperbolic motion.

Away from a linear neighborhood of $E$, the linear approximation is no longer
valid. Higher order terms become important for the dynamics, and the
non-in\-te\-gra\-bi\-li\-ty of the system results in the distortion and twisting of the
cylinders. The stable cylinder and the unstable one will cross each other
transversally, producing chaotic sets\cite{ivano}\,\cite{wiggins} in the phase
space; orbits with initial conditions taken on these chaotic sets will be
highly sensitive to small perturbations on the initial conditions.

The phase space under consideration is not compact, and we will actually
identify chaotic behaviour associated to the possible asymptotic ($t\rightarrow \infty$) outcomes of orbits, namely, escape to the de Sitter attractor at
infinity or collapse after a burst of initial expansion. We assume
$\Lambda=0.25$, so that the critical point $E$ is characterized by $A=B=1.0$,
$P_A=P_B=0$ and $E_{crit}=1.0$. All calculations were made using the package
$Poincare$\cite{ed}. Let $S_0$ be a point belonging to the separatrix
($E_0=1.0$), with coordinates $A=B=0.4$, $P_B=2\,P_A=1.3576450198$. Around this
point, we construct a 4-dim sphere in the phase space with arbitrary small
radius $R_0$, for instance $R_0=10^{-5}$. The values of $(A,B,P_A,P_B)$ are
taken in energy surfaces which have a non-empty intersection with this sphere.
Such energy surfaces are those for which the range of energy, $E_0$, about
$E_0=E_{crit}=1.0$ is of the order of, or smaller than the radius $R_0$.
Physically we are considering expanding cosmological models with small
anisotropic perturbations in a pre-inflationary phase.

After several experiments with the above set of initial conditions, we note
that, as expected, two possible outcomes arise: collapse, or expansion into
the de Sitter configuration. The main result of this paper is to show that,
for a determined interval of energy $\delta\,E*$, this outcome is chaotic.
Indeed, according to our numerical work, we find out that, for each sphere of
initial conditions, there always exists a non-null interval $\delta\,E^*$ for
which orbits either collapse or escape, in an indeterminate outcome. In Fig. 3
this behaviour is showed for spheres of initial conditions with radius
$R_0=10^{-5}$. An empirical relation between the gap $\delta\,E^*$ and the
radius $R_0$ is obtained, $\delta\,E^* \propto R_0^2$. In other words, any
infinitesimal fluctuation of a given initial condition in the sphere (for
energies in the interval $\delta\,E^*$ can lead to an indeterminate outcome, that is,
collapse or escape. This is evidence of chaos, as a consequence of the
crossing of stable and unstable cylinders, emanating from unstable periodic
orbits about $E$. This topological structure is actually an invariant
characterization of chaos. Finally we remark that the above behaviour is not
restricted to initial conditions taken in small neighborhoods of points of the
separatrix. Generally, any sets of initial conditions taken in an arbitrary
neighborhood of the invariant manifold ${\cal{M}}$, which result in orbits
that visit a small neighborhood of $E$, display the above chaotic behaviour.

\begin{figure}[htb]
\centerline{
	\hbox{
        \psfig{figure=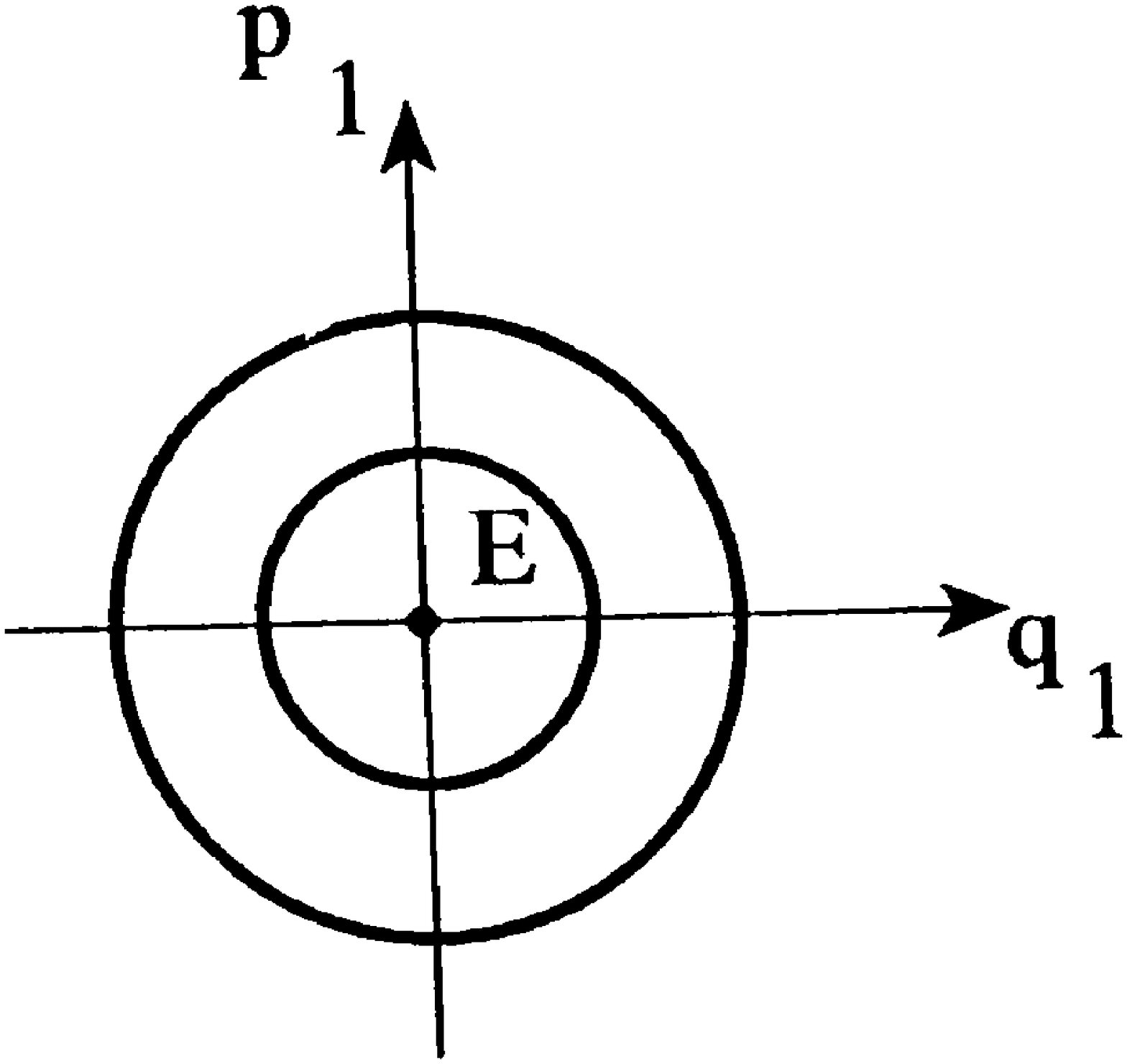,width=35mm,height=40mm}
           \hspace{10mm}      
         \psfig{figure=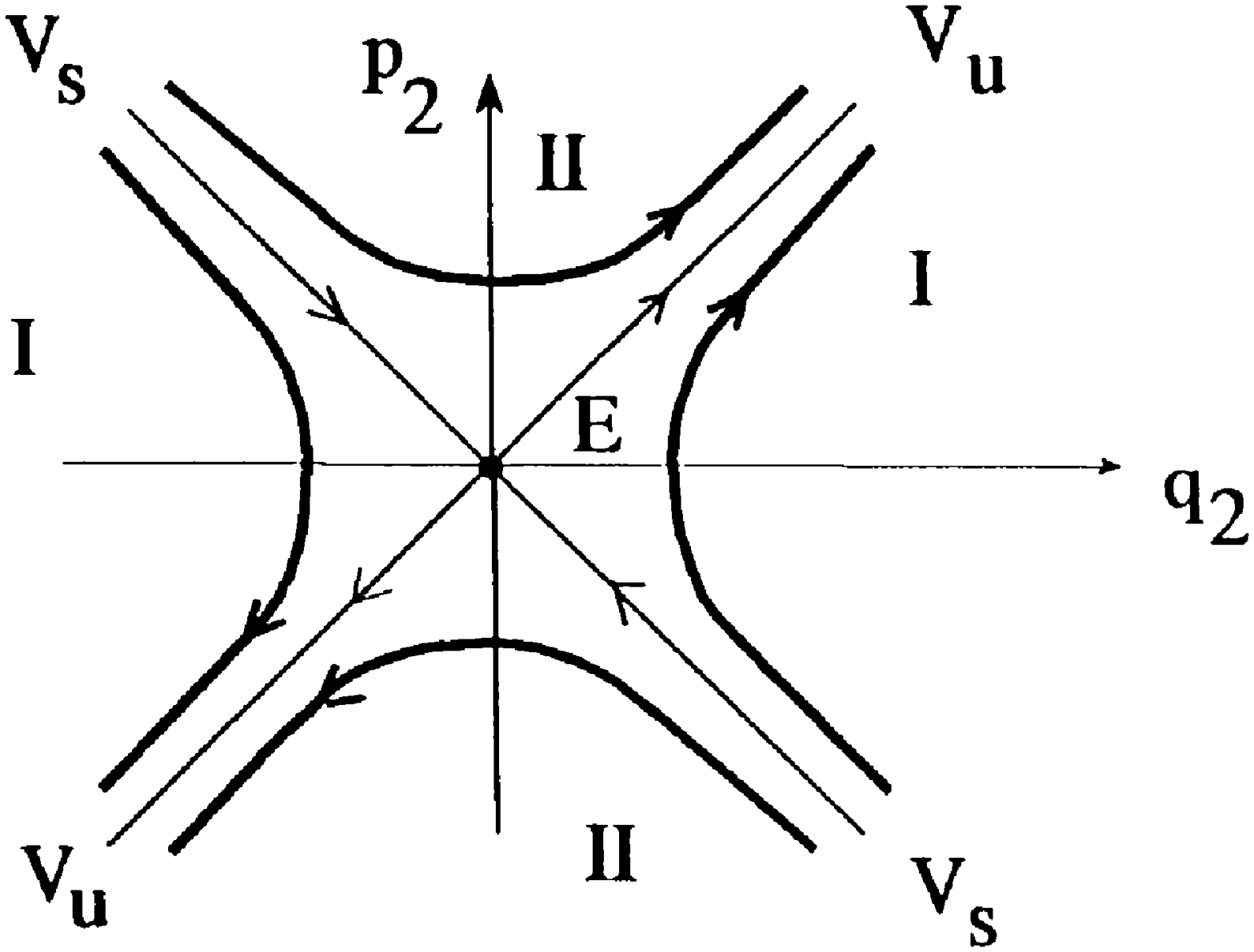,width=30mm,height=30mm}\hspace{10mm}
        \psfig{figure=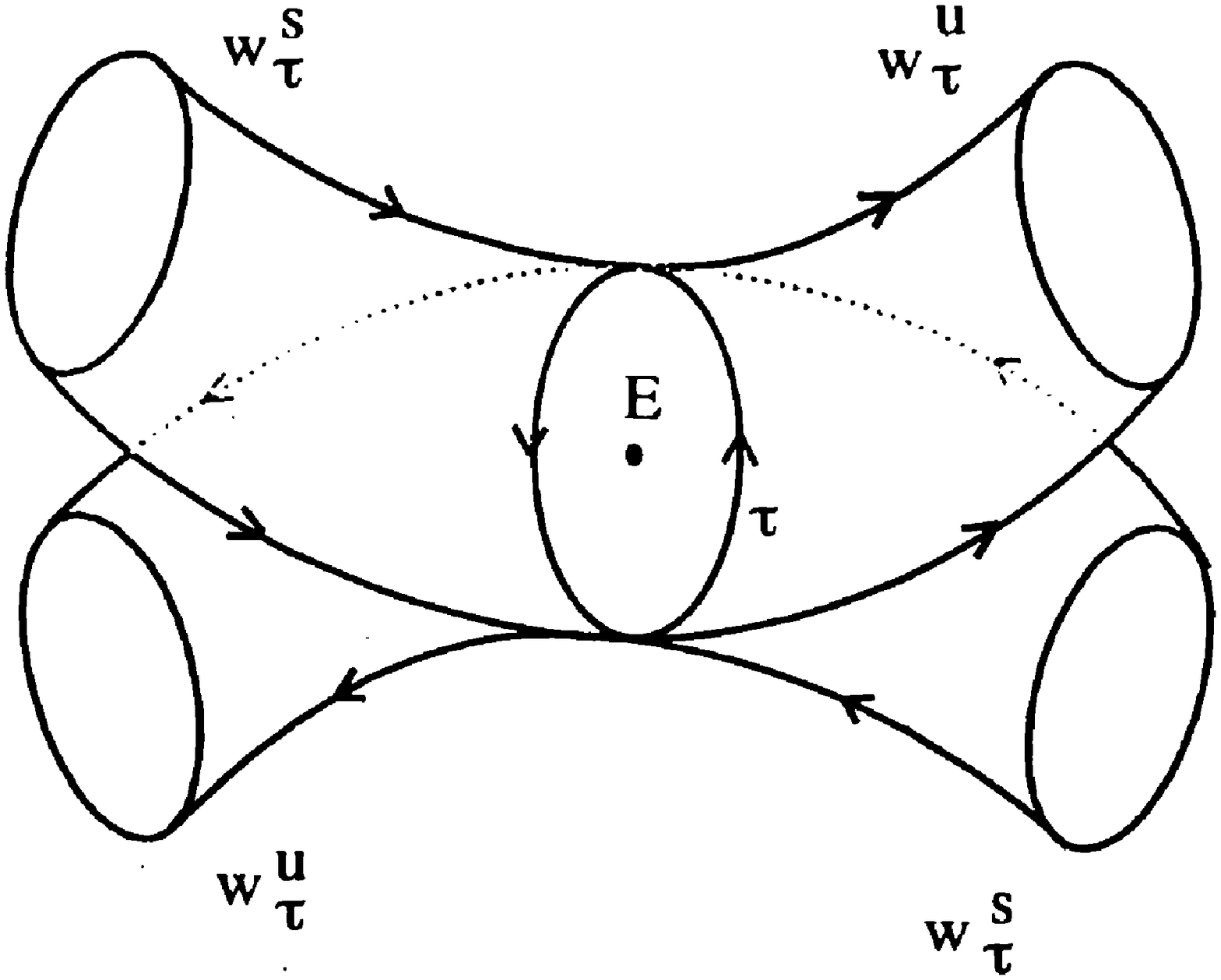,width=35mm,height=35mm}    
	}	
}
	\caption{(a)Periodic orbits of the Hamiltonian in teh linear approximation.
	(b)The linear unstable $V_u$ and stable $V_s$ one-dimensional manifolds. (c)Stable and unstable cylinders manifolds emanating from the periodic orbit $\tau$.}
\end{figure}

\begin{figure}[htb]
\begin{minipage}[t]{0.49\textwidth}
\epsfysize=4cm
\epsfxsize=5cm
\epsffile{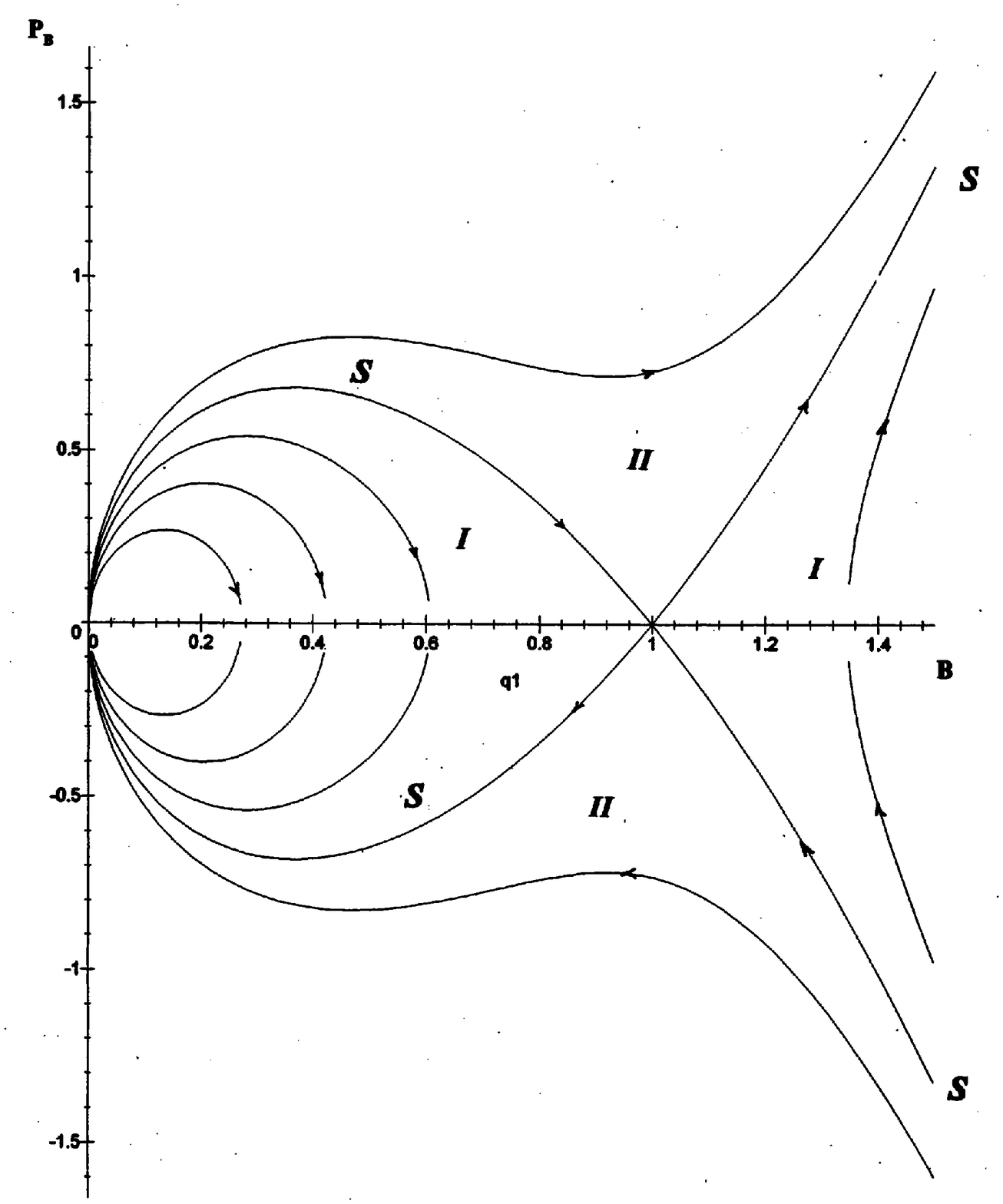}
\caption{Integral curves on the invariant manifold $A=B$, $P_A=P_B/2$ for $\gamma=1$ (dust).}
\end{minipage}
\hspace{0cm}
\begin{minipage}[t]{0.49\textwidth}
\epsfysize=4cm
\epsfxsize=4.5cm
\epsffile{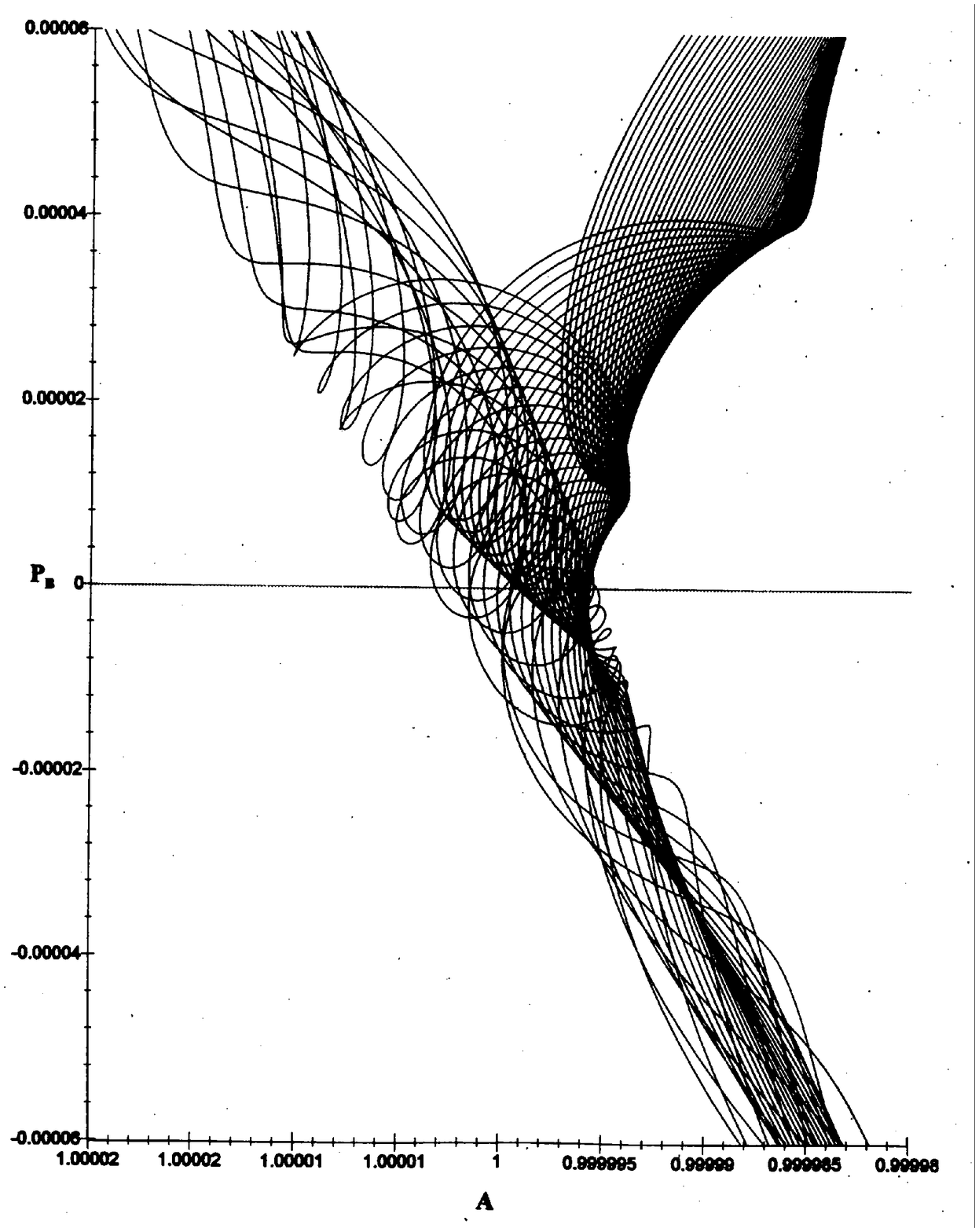}
\caption{Chaotic exit to inflation: outcome of 50 orbits choosen in a sphere of $R=10^{-5}$ about a point $S_0$ on the separatrix.}
\end{minipage}
\end{figure}
%

%\begin{figure}[htb]
%\centerline{
%	\hbox{
%         \psfig{figure=fig1.eps,width=30mm,height=30mm}\hspace{15mm}  
%         \psfig{figure=fig3.eps,width=35mm,height=30mm}}}
%	\caption{(a) bla bla.
%	(b) bla bla.}
%\end{figure}

\vspace{-3mm}

\section*{Acknowledgments}The authors are grateful to CNPq and FAPERJ for
financial support.

\end{document}